\documentclass[pre,twocolumn,showpacs,aps]{revtex4}

\usepackage[english]{babel}
\usepackage{amssymb,amsfonts,amsmath}
\usepackage[pdftex]{graphicx}
\usepackage{multirow}
\usepackage{color}

\newcommand{\mc}{\mathcal{C}}
\newcommand{\mb}{\mathcal{B}}

\begin{document}

\title{Statistical significance of communities in networks}

\author{Andrea Lancichinetti}\affiliation{Complex Networks Lagrange Laboratory (CNLL), ISI Foundation, Turin, Italy}\affiliation{Physics Department, Politecnico di Torino, Turin, Italy}
\author{Filippo Radicchi}\affiliation{Complex Networks Lagrange Laboratory (CNLL), ISI Foundation, Turin, Italy}
\author{Jos\'e J. Ramasco}\affiliation{Complex Networks Lagrange Laboratory (CNLL), ISI Foundation, Turin, Italy}

\widetext

\begin{abstract}
Nodes in real-world networks are usually organized in local modules. These groups, called communities, are intuitively defined as sub-graphs with a larger density of internal connections than of external links. In this work,
we introduce a new measure aimed at quantifying the statistical significance of single communities. Extreme and Order Statistics are used to predict the statistics associated with individual clusters in random graphs. These distributions allows us to define one community significance as the probability that a generic clustering algorithm finds such a group in a random graph. The method is successfully applied in the case of real-world networks for the evaluation of the significance of their communities.
\end{abstract}


\maketitle 

\section{Introduction}

Complex networks play a crucial role in understanding physical, biological, social and technological systems~\cite{albert02,newman03,romu_alex04}. Interactions between proteins in cells of living organisms, relations between human actors in socioeconomic contexts and connections between Web pages in the World Wide Web can naturally be described as graphs. Real-world networks typically have complex topological properties, but in spite of their evident diversity, structural analysis has revealed that they share a conspicuous set of common features: {\it scale-free}ness ({\it i.e.}, the number of connections per node following a wide or  power-law distribution)~\cite{albert02} and {\it small-world}ness ({\it i.e.}, the average number of hops between two nodes in the network scales logarithmically with its size)~\cite{watts98} are two celebrated examples of such properties. Recent studies have focused on deeper structural features of networks. Real-world networks are typically organized in local clusters of nodes which are usually denominated communities. Communities are groups of nodes with a higher level of interconnection among themselves than with the rest of the graph. In this sense, communities are groups relatively isolated from the other nodes of the network and are expected to represent elements sharing common features and/or playing similar roles within the system (see Ref.~\cite{santo09} for an exhaustive review). For instance, if one considers the World Wide Web, communities are composed by groups of Web pages dealing with similar topics; in social networks, communities stand for sets of actors sharing common interests, ideas and friendship relationships; in protein interaction networks, communities represent groups of proteins with similar functionalities.

This imbalance of in- and out-connections corresponds to an intuitive concept. There are some formalizations of the definition of community. The {\it LS set}~\cite{seidman83} or {\it strong community}~\cite{girvan02,radicchi04} stands for a group where every node belonging to the group has more internal connections than external ones. A less restrictive definition refers to a {\it weak community}~\cite{radicchi04} as a set of nodes where the number of intracommunity connections (summed over all nodes within the group) is larger than the number of links going out of the community. Along these lines, the well known {\it modularity}  is a quality function able to quantify the statistical importance of a partition comparing the number of internal connections observed in the communities with its expected number in a suitable {\it null model}~\cite{newmang04}. Besides the formulation of a definition, big efforts have been made for the detection of communities in networks. Since the total number of possible divisions of a network in subgraphs is a non-polynomial function of the size of the network itself, finding and detecting communities is not a trivial issue. Many algorithms have been proposed during recent years, every of them with the same spirit of finding the best groups which maximize the internal density of links~\cite{santo09,newmang04,mcl00,eriksen03,zhou03,reichardt04,newman04,donetti04,duch05,palla05,newman06,arenas06,sales07,kumpula07,newman07,ramasco08,rosvall08}. Different principles may be used, but nevertheless in all cases some property related to the community structure is locally or globally optimized. The consequence is that even in uncorrelated networks these algorithms find clusters that are supposed to be good according to the modularity function or to other quality measures. 

If algorithms are able to identify communities even in random graphs, which value can we give to communities found in real networks? Or better, how to statistically determine the significance of a community? This problem has been the subject of some studies in the literature~\cite{sales07,spirin03,guimera04,reichardt06,reichardt08,karrer08,bianconi09}. In~\cite{sales07,reichardt06} for example, the partition of a network maximizing the modularity is compared with the maximum modularity partition of a randomized version of the given network ({\it i.e.}, all edges are randomly rewired). In~\cite{karrer08}, differently, the importance of a community partition is proportional to its robustness against random perturbations ({\it i.e.}, random reshuffling of edges). Such heuristic approaches rely on the modularity function to evaluate the quality of a partition, which means that are subjected to the modularity resolution limits~\cite{kumpula07,fortunato07}. Furthermore, all the proposed methods are designed to deal with full partitions, not with single communities. Even though in a network one might find some meaningful communities alongside with randomly connected node clusters. In this paper, we develop a statistical method aimed at discriminating between a single {\it bona fide} community and structures arising as topological fluctuations. Instead of a direct comparison with an average outcome, the community is confronted with the best expected result for a null-model. The reason for stressing this "best outcome" is that community detection algorithms will in general produce the best possible clusters given a graph, even if it is random. The threshold of significance can be approximated by using Extreme and Order Statistics~\cite{david03,beirlant04} applied to null-model community fitness. A community significance can be then obtained as the extreme probability of finding a group equal or better than the one given in a set of equivalent random graphs.

\section{Null-models and definition of \textit{c}-score}

\begin{figure}[t]
\begin{center}
\includegraphics[width=8.6cm]{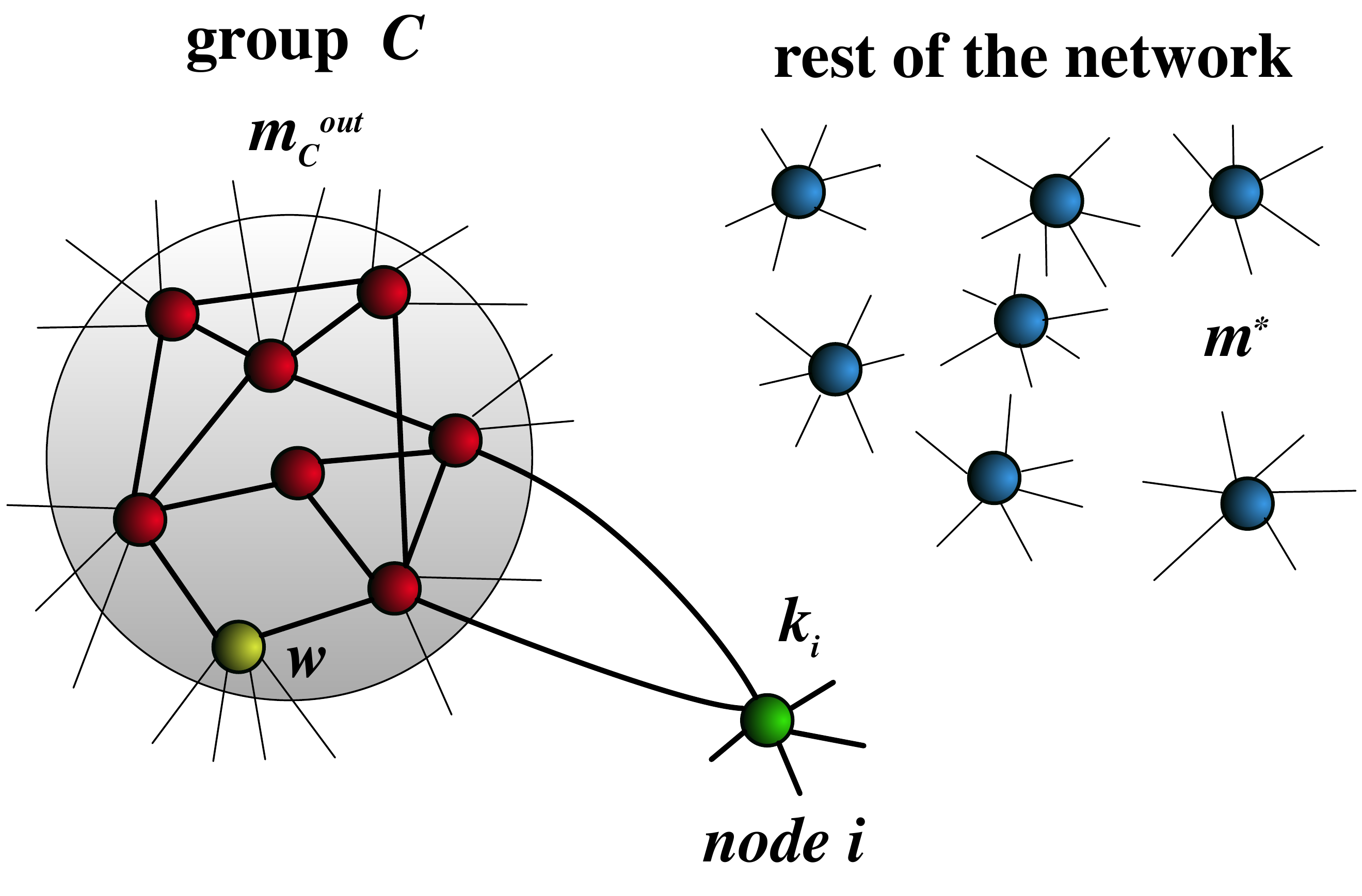}
\caption{(Color online) Sketch of the theoretical framework referring to the null-model. Node $i$ has $k_i$ free ends to allocate. Each of them can connect to nodes within $\mc$ or  vertices belonging to the rest of the network.\label{fig:sketch}}
\end{center}
\end{figure}

Consider a scenario as the one depicted in Fig.~\ref{fig:sketch}, with a given community $\mc$ in a graph. $k_i$ denotes the number of connections (degree) of the node $i$. Given $\mc$, $k_i$ can be divided in two terms: $k^{int}_i$ , the number of links connecting $i$ to nodes in $\mc$, and $k^{ext}_i$, the number of connections outside. Similarly, we define the internal degree of $\mc$, $m^{int}_\mc =\sum_{j\in \mc} k^{int}_j$, as well as $m^{ext}_\mc =\sum_{j\in\mc} k^{ext}_j$ and its total degree $m_\mc = m^{int}_\mc +m^{ext}_\mc$. We consider a very simple stochastic null model: all connections inside the group are locked (the community is given so cannot be altered), while the other links are randomly reshuffled among all nodes preserving their degrees. For simplicity, we allow the rewiring operation to form multiple links (two nodes can be connected by more than one edge) or self-loops. In some weighted graphs the weights of the links are equivalent to multiple connections and so the present null-model would be appropriate. Some examples are social networks (the Zachary club~\cite{zachary77}, see last section) or the {\it C. Elegans} metabolic network~\cite{duch05} that will be analyzed later. For unweighted graphs, we have checked that our results do not noticeably change by including or not multiple links as long as the graph is not condensed (a node gets a finite fraction of the links). When node degrees are much smaller than the network size, the probability of generating self-loops and multiple links by random reshuffling becomes negligible. Note also that our null model is similar to the one used for the definition of the modularity~\cite{newmang04} and close in spirit to the configurational model~\cite{molloy98}. It generates graphs that have no special internal structure except that given by random fluctuations, keep the degree sequence of the original network and can show degree-degree correlations only if the degree sequence and the network size determine their presence~\cite{catanzaro05}. This is the most general null model, appropriate when no knowledge about the system is available and simple enough to be treated from an analytical point of view. If further information regarding the constraints present in the process that generated the given network is available, other,  simpler or more elaborated, null models can be employed. Our method to evaluate group significance is general enough to admit the use of different null models by altering consequently the distributions that will be described  next. 

\begin{figure}
\begin{center}
\includegraphics[width=8.6cm]{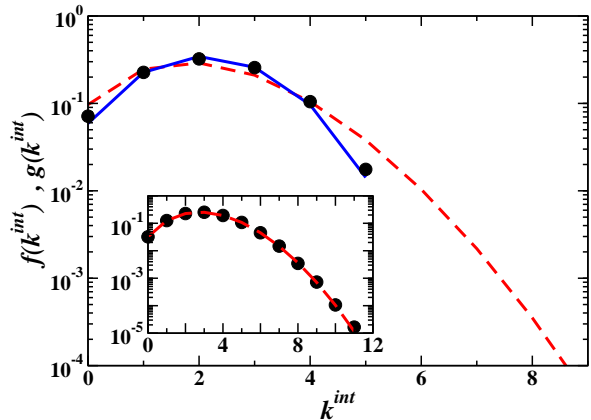}
\caption{(Color online) Distributions $f(k^{int})$ and $g(k^{int})$ for randomly generated networks of homogeneous degree. The (black) circles are numerical results with $N = 100$, $n_C = 20$ and $k = 15$. The distributions refer to external nodes of groups selected at random (inset) or by maximizing modularity (main plot). Dashed (red) curves are the approximation given by Eq.~(\ref{hypergeom}) and the continuous (blue) one that of Eq.~(\ref{muht}).\label{fig:hyper}}
\end{center}
\end{figure}

Once the null model has been selected, suppose that $\mc$ is a group composed of randomly chosen nodes and consider a generic node $i$ not belonging to $\mc$. The distribution of $k_i^{int}$ is given by the hypergeometric distribution
\begin{equation}
\label{hypergeom}
f(k_i^{int} \mid \mc) =  \frac{ {m^{ext}_{\mc}\choose k_i^{int}}  \; \times\; {m^*  - m^{ext}_{\mc} \choose k_i - k_i^{int}} }{ {m^* \choose k_i}} \;\;,
\end{equation}
where ${y \choose x} =\frac{y!}{\left(y-x\right)!x!}$ is a binomial coefficient, and $m^*$ are the free ends in the network: $m^* = m - m_\mc$ ($m$ are the total ends in the graph, twice the number of links). Eq.~(\ref{hypergeom}) states that the probability of node $i$ to have $k_i^{int}$ internal connections to $\mc$ is given by the ratio of two terms: the total number of ways in which $k_i^{int}$ links can be placed at the end of $m^{ext}_{\mc}$ free ends multiplied by the number of ways to locate the remaining $k_i - k_i^{int}$ edges out of $m^* - m^{ext}_{\mc}$ free ends, divided by the total number of ways to place all $k_i$ connections in the network (i.e., out of $m^*$ free ends). If the node $i$ belongs to $\mc$, Eq.~(\ref{hypergeom}) has to be corrected to exclude $i$ from the group. When the group $\mc$ is composed of $n_\mc$ randomly chosen nodes, Eq.~(\ref{hypergeom}) recovers the results obtained via numerical simulations (see inset Fig.~\ref{fig:hyper}). 

The next, more interesting, case is when $\mc$ is not composed of randomly chosen nodes, but it has been detected by a clustering algorithm. As can be seen in the main plot of Figure~\ref{fig:hyper}, the shape of $f(k^{int})$ dramatically changes due to the algorithm node selection. Most of the nodes populating the tail of the distribution are incorporated into the group. Correlations are also present since nodes in the community are expected to be connected among themselves. Still, it is possible to obtain an approximate expression for the probability $f(k^{int})$. We first consider the case of homogeneous graphs where all nodes have the same degree (i.e., $k_i=k \;, \, \forall \, i$) and extend later our analysis to networks with arbitrary degree sequences. We will assume that $\mc$ has been selected to maximize $k^{int}_{i}$ for each node inside $\mc$ as well as the overall $m^{int}_{\mc}$. This also implies that since the nodes are all equivalent, have the same degree $k$,  they can be ranked according to their $k^{int}$. We indicate with $w$ the node (or nodes) with the lowest $k^{int}$ within the community (see Fig.~\ref{fig:sketch}). $k_w^{int}$, the internal-degree of the worst node,  establishes then an upper cut-off to the possible values of $k^{int}$ of the out-group nodes. An expression similar to Eq.~(\ref{hypergeom}) can then be derived for the external nodes by taking into account this new cut-off
\begin{equation}
\label{muht}
g(k_i^{int}\mid\mc, k^{int}_w ) =  \frac{ {m^{ext}_{\mc}\choose k_i^{int}}  \; \times\; {\widetilde{m}  - m^{ext}_{\mc} \choose k^{int}_w - k_i^{int}} }{ {\widetilde{m} \choose k^{int}_w}} \;\;,
\end{equation}
where $\widetilde{m} = (N-n_{\mc}) \,k^{int}_w$. The term $\widetilde{m}$ accounts for the fact that no node can connect to more than $k^{int}_w$ internal vertices and therefore some of the free ends $m^*$ become occupied.  Eqs.~(\ref{hypergeom}) and  (\ref{muht}) specify the null-model. Our method does not depend on the particular functional shapes of $f(k_i^{int})$ and $g(k_i^{int})$. For instance, a more restricted null-model without multiple links can be approximated by using Wallenius hypergeometric distribution, although this considerably complicates the numerical evaluation of the functions. Another null-model, less realistic but very easy to implement, is the Erd\"os-R\'enyi-like networks for which $f(k_i^{int})$ and $g(k_i^{int})$ are binomial distributions.

\begin{figure}
\begin{center}
\includegraphics[width=8.6cm]{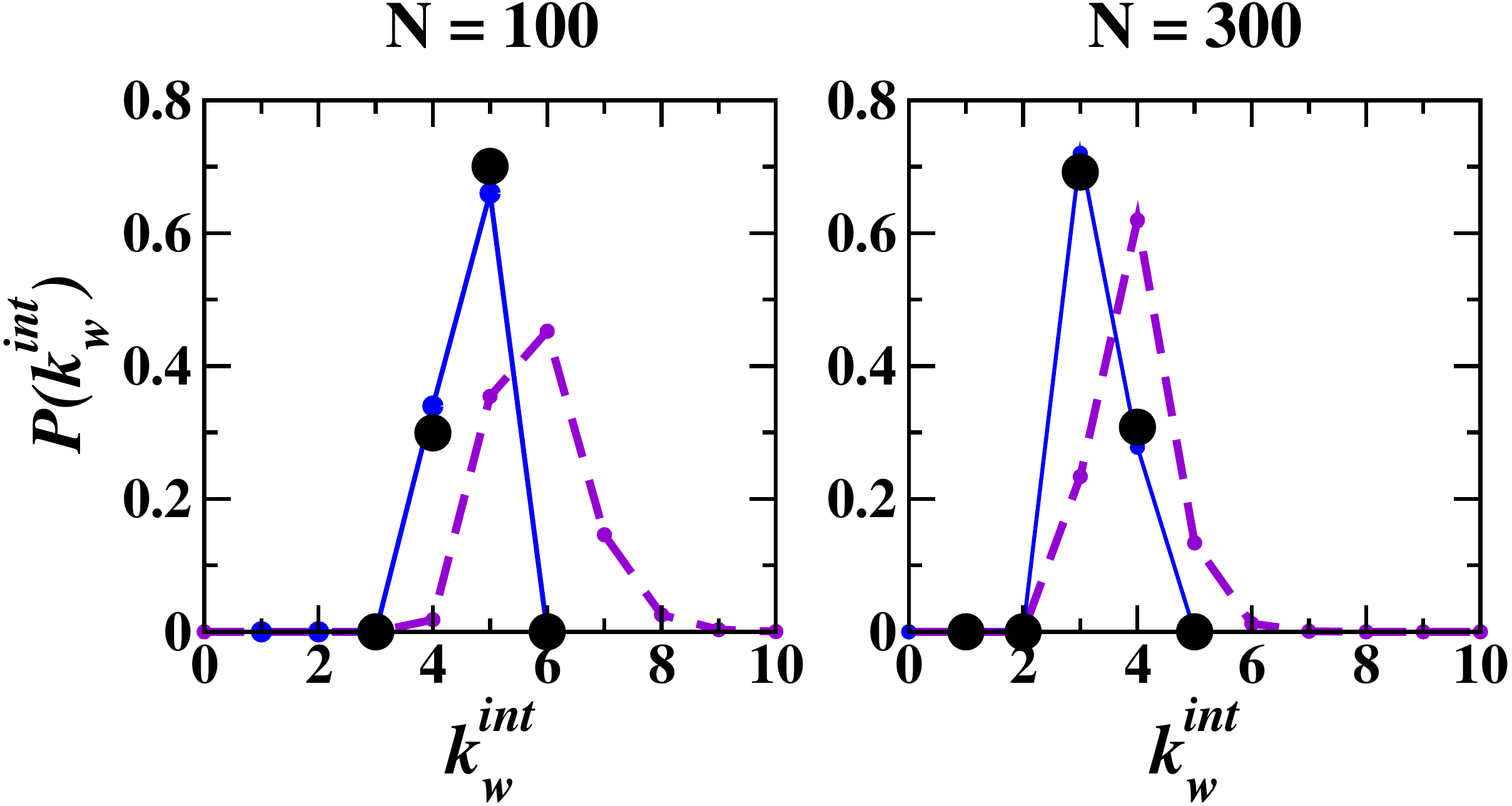}
\caption{(Color online) Probability distribution $P(k^{int}_w)$ for the internal degree of the worst node of the community calculated for groups $\mc$ detected by maximizing the modularity in randomly generated networks of $N=100$ (left) and $N = 300$ (right) with homogeneous degree $k=15$ and for groups of $n_\mc = 20$. The black circles are the results of numerical simulation, while the continous blue curves correspond to the theoretical predictions derived from Eq.~(\ref{pkw}). The extreme value distribution of $f(k^{int})$ from Eq.~(\ref{hypergeom}) is also plotted for comparison (violet dashed curve).\label{fig:kw}}
\end{center}
\end{figure}

The worst node within the community, $w$, will play a central role in our method
to evaluate group significance. We assume that in a random graph there is not a drastic variation between $k^{int}_w$ and the internal degree $k^{int}$ of the best nodes outside the group. Postulating a smooth variation of $k^{int}$ between inside and outside of the community allows us to find an expression for the probability distribution of $k^{int}_w$ based on Eq.~(\ref{muht}) which only applies to external nodes. The degree of the worst node, $k^{int}_w$, is a given quantity in $g(k^{int}\mid \mc, k^{int}_w)$. In order to find a formula for $P(k^{int}_w)$, we need thus to alter our point of reference and consider the second worst node within the community $w'$. If the statistics of $k^{int}_w$ is comparable to that of the best external nodes, $k^{int}_w$ should follow the distribution of the extreme of $g(k_i^{int}\mid \mc\setminus\{w\}, k^{int}_{w'})$. This means that the probability for $k^{int}_w$ to be lower or equal to a certain number reads
\begin{equation}
\label{pkw}
\mbox{Pr}(\le k^{int}_w ) =  [G( k^{int}_w |\, \mc \setminus \{w\}, k^{int}_{w'})]^{N-n_\mc+1} \;\;,
\end{equation}
where $G(\cdot)$ is the cumulative of the function $g(\cdot)$. The distribution $P(k^{int}_w)$ is given by the derivative of the cumulative of Eq.~(\ref{pkw}), $P(k^{int}_w) = \partial \, \mbox{Pr}(\le k^{int}_w )$. It must be remarked that Eq.~(\ref{pkw}) is valid for independent random variables, in our null model the independence is justified for external nodes and is an approximation when refers to $w$. Figure~\ref{fig:kw} shows a comparison between the distribution $P(k^{int}_w)$ obtained with this procedure and its counterpart from numerical simulations.  Despite the approximations performed to reach an analytical form for $P(k^{int}_w)$, the agreement is remarkable. The use of Extreme Statistics contributes in part to such agreement, since under very general conditions the limit extreme-value distribution is stable and has no memory of the parental distribution.

Once a functional form for $\mbox{Pr}(\le k^{int}_w )$ was obtained, we can define a measure of the significance of a group, the $\mc$-score, as 
\begin{equation}
\label{cscore}
c= \mbox{Pr}(\ge k^{in}_w ) = 1 - \mbox{Pr}(\le (k^{int}_w-1) ) ,
\end{equation}
which corresponds to the probability that $k^{int}_w$ for an optimized community in an equivalent random graph ensemble is higher than or equal to the value seen in $\mc$. A point to stress here is that $c$ contains not only information about the worst node, $k^{int}_w$, but also about the community external links and about the degrees of the external nodes.

In order to extend our results to heterogeneous graphs, we need to rank the nodes according to the role they play with respect to the given community $\mc$. For regular networks, since all the nodes are equivalent, the ranking can be simply established by considering the values of the internal degrees $k^{int}$. However,  another criterion is required to deal with heterogeneous networks. We use the probability distribution provided by Eq.~(\ref{hypergeom}) as the basis for such procedure. The rank for a node $i$ can be established by the probability of finding a node with an internal degree $k_i^{int}$ or higher in the null model given its degree $k_i$ and $\mc$. That is, for each node $i$ we calculate the score $r_i = \sum_{q=k^{int}_i}^{k_i} f\left(q\right)$ and then perform comparisons on the basis of $r$. The values of $r$ fall in the interval $\left[0,1\right]$ regardless of the node degree, which facilitate the comparison. $w$ and $w'$ correspond thus to the nodes with the highest and second highest values of $r$ within the community, respectively. Under the hypothesis of a randomly connected network, the scores $r$ of the vertex $w$, $r_w$, and that of the external nodes can be seen as random variables uniformly distributed in the interval $[r_{w'}, 1]$. The $\mc$-score can be then calculated as the probability of observing $r_w$ as the minimal value of a set of $(N-n_C+1)$ random extractions from a uniform distribution defined in the interval $[r_{w'}, 1]$. An alternative to this last step is to map the internal degree of $w'$ into $\widehat{k}^{int}_{w'}$ (the internal degree that it would have if its degree was equal to $k_w$ and its score $r_{w'}$) by inverting the distribution of Eq.~(\ref{hypergeom}). Once the transformation has been performed, we can proceed in the same way as for homogeneous networks with Eqs.~(\ref{pkw}) and (\ref{cscore}).

\section{Beyond the $\mc$-score} 
\label{bb}
A low value of the $\mc$-score ({\it i.e.}, $c \leq 5\%$) is enough to consider a group as significant. However, when the $\mc$-score is higher, one could argue that the reason is that relaying only on the worst node of the community for the full group evaluation is a too severe criterion. Algorithms may fail to place a single node and this would translate into a non significant community according to the $\mc$-score approach.  The performance of the method can be improved by a further refinement. Instead of considering only the last node, one can include a longer list of nodes and use this information for the computation of the statistical significance of the community. A way to do so is to write an algorithmic procedure.  Three classes of nodes can be considered: The community $\mc$, the "border" $\mb$ and the rest of the network. Initially, the group $\mb_0$ is empty and $\mc_0 = \mc$. Then at each algorithm step, the following actions are taken
\begin{itemize}
\item{Compute  $r_i = \sum_{q=k_i^{int}}^{k_i} \, f(q)$, where the function $f(\cdot)$ is given by Eq.~(\ref{hypergeom}). $r_i$ is calculated for each node $i \in \mc$ with respect to the group $\mc_{t}$;}

\item{Determine the worst node in $\mc_t$,  $w_{t+1}$, as the vertex with highest
 $r_{w_{t+1}}$. Set $\mb_{t+1}=\mb_{t} \cup \{w_{t+1}\}$ and $\mc_{t+1}=\mc_{t} \setminus \{w_{t+1}\}$;}
\item{Compute $\mbox{Pr}(< S_{t+1} | \mc_{t+1}, \mb_{t+1}, r_{w_{t+2}} )$, where $S_{t+1} = \sum_{i \in \mb_{t+1}} r_{w_i}$ and $w_{t+2}$ is the worst node still in $\mc_{t+1}$ ;}
\item{Increase $t \to t+1$.}
\end{itemize}  
This algorithm explores the interior of the community trying to maintain the worst nodes always in $\mb$, it ends when $t=n_{\mc}-1$. $\mbox{Pr}( < S_{t+1} | \mc_{t+1}, \mb_{t+1}, r_{w_{t+2}} )$ 
stands for the probability that the sum of the scores of the worst $t$ nodes of an optimized community in an ensemble of equivalent random graphs  is smaller than the given for $\mc$. Its value for a set of independent random variables can be estimated by using Order Statistics (see Appendix~\ref{app} for more details). We define then the $\mb$-score as
\begin{equation}
\label{bscore}
\mb\mbox{-score}  = \min_t \; \mbox{Pr}( < S_t | \mc_t, \mb_t, r_{w_{t+1}})\, ,
\end{equation}
which corresponds to the lowest value of the probability $\mbox{Pr}( < S_t | \mc_t, \mb_t, r_{w_{t+1}})$ observed during the iterative procedure. We take the minimum as the best approximation for the significance of the group $\mc$, since it is evaluated in the most favorable discrimination of $\mc$ nodes in border and core. This probability is equivalent to the $\mc$-score for $t = 1$, while becomes a more synergic quantity as $t$ increases. The inclusion of a longer list of worst nodes in the calculation helps to correct conservative estimates due to under-sampling. When communities are significant with respect to the $\mc$-score they are significant also according to the $\mb$-score. {\it Vice versa}, low values of the $\mb$-score do not necessarily correspond to small $\mc$-scores. Many concomitant bad nodes with features slightly different from the random expectations may multiply their effect and lead, if there is a real signal, to the prediction of a significant community.

\section{Computational benchmarks}

\begin{figure}
\begin{center}
\includegraphics[width=8.5cm]{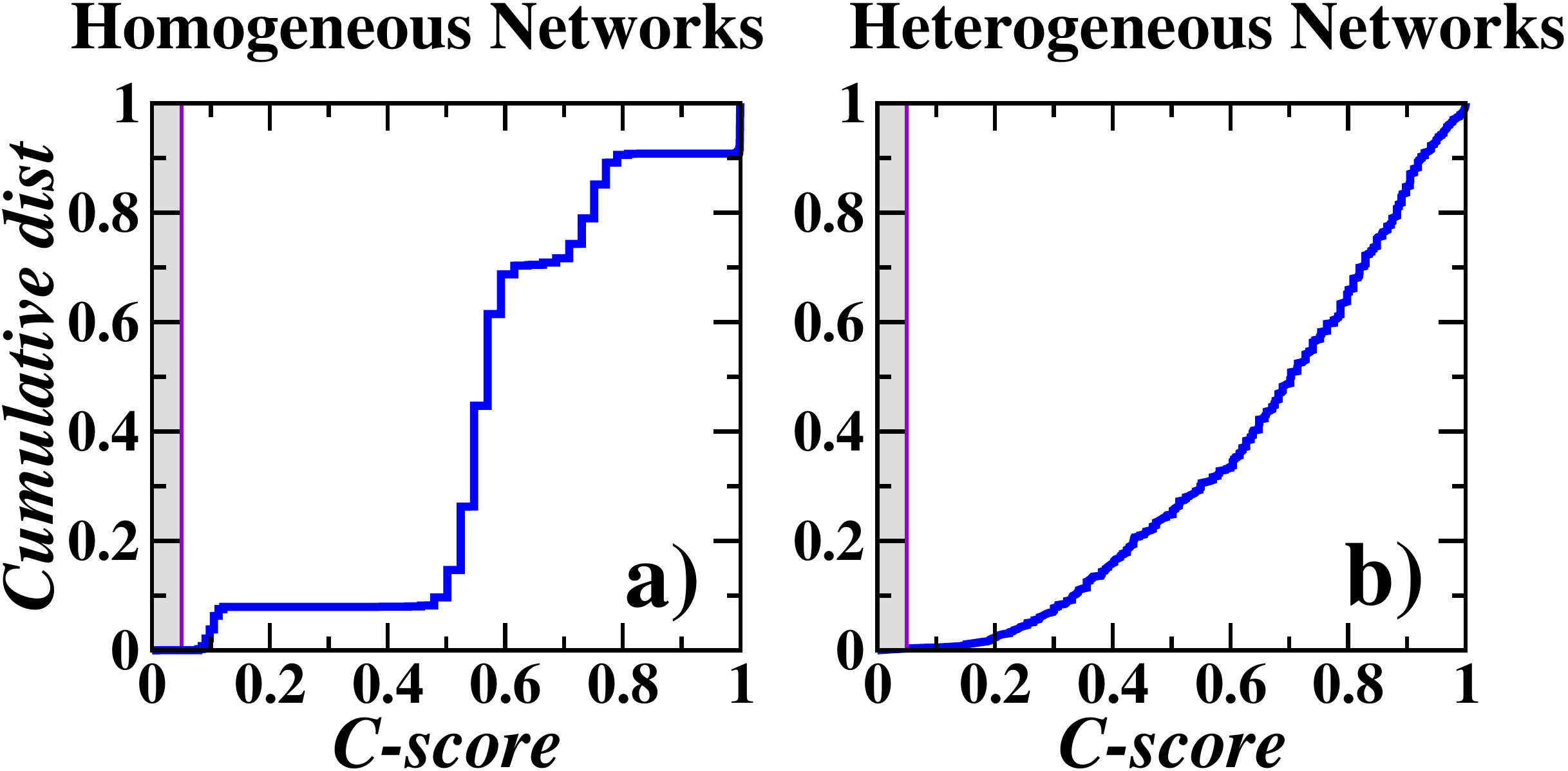}
\caption{(Color online) Cumulative distribution of the $\mc$-score for groups of size $n_\mc=20$ obtained in random networks of $N = 100$ nodes by modularity maximization. (a) Homogeneous networks with degree $k=15$. (b) Heterogeneous networks with average degree $\langle k \rangle =15$. The degree distribution follows a power-law $P(k) \sim k^{-\gamma}$ with $\gamma=2$. The gray areas delimit the values of $\mc$-score that indicate group statistical significance (probability $\le 5 \%$).}
\label{fig:ran_test}
\end{center}
\end{figure}

As a first test, we applied the $\mc$- and the $\mb$-scores to groups found in random graphs using clustering techniques. The $\mc$-score and the $\mb$-score are able to identify these groups as not significant (see Figure~\ref{fig:ran_test}). The results confirmed that the scores are good estimators for the statistics of such groups further contributing to our confidence in the method. We consider next  the performance of the scores on artificial networks with planted community structure. In order to do so, we build networks in the spirit of Girvan and Newman's benchmark~\cite{girvan02}. Since our aim is to evaluate a single cluster, the benchmark will be composed of a group $\mc$ with $32$ nodes and of other $96$ nodes in the rest of the network. Every node in $\mc$ is connected on average with $\langle k^{int} \rangle$ nodes of its own group and $\langle k^{ext} \rangle$ outside. The external nodes are connected at random. The average total degree for all the nodes is fixed at $\langle k \rangle =16$. $\langle k^{ext} \rangle$ acts thus as a control parameter for the strength of the community structure. The higher it is, the more 
\begin{figure}[b]
\begin{center}
\includegraphics[width=8.5cm]{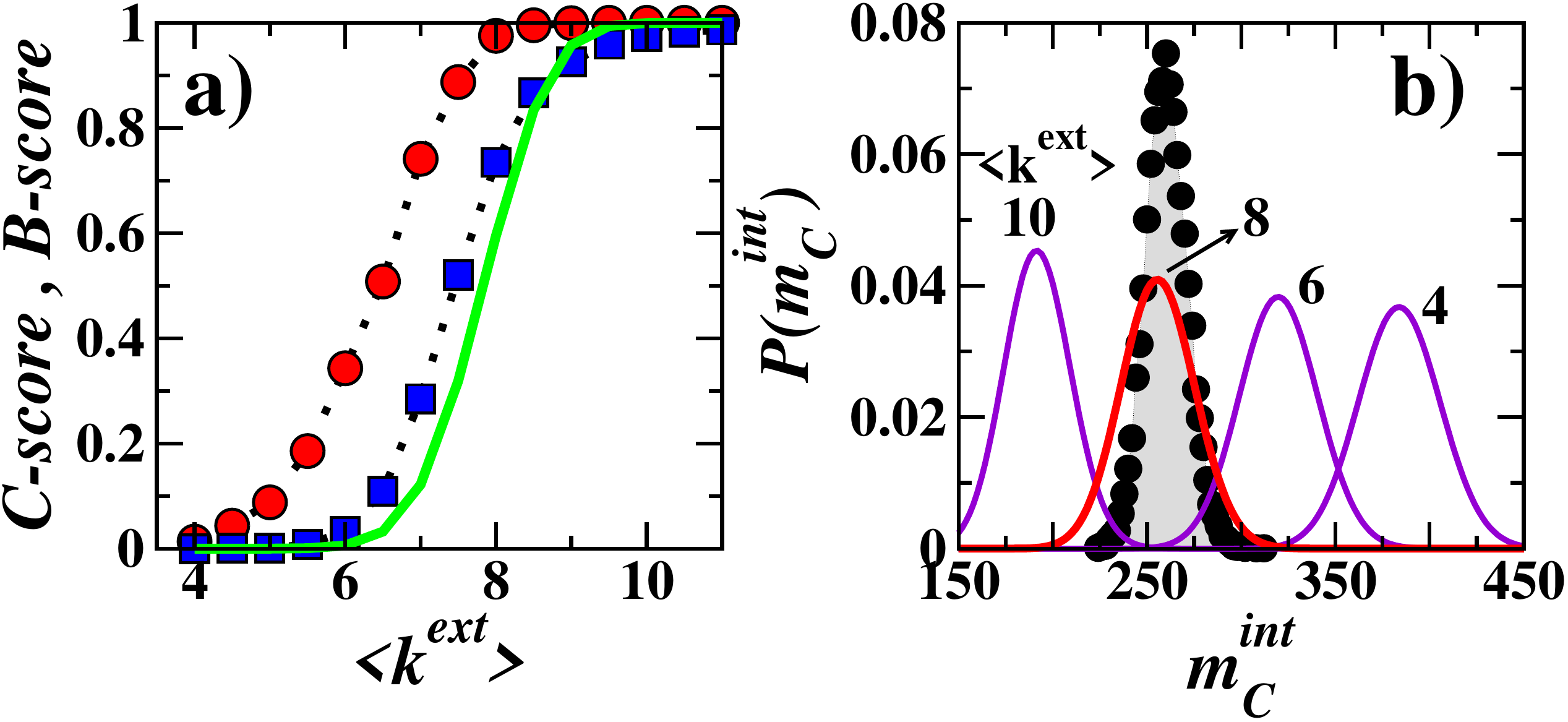}
\caption{(Color online) a) $\mc$- and $\mb$-scores for communities in benchmarks. The disorder in the connections increases with $\langle k^{ext}\rangle$. The continuous (green) curve corresponds to the target distribution obtained numerically (see text for details). b) Distribution $P(m^{int}_\mc)$, continuous curves are for the benchmark with the $\langle k^{ext}\rangle$ shown over the curve. The black circles are the numerical distribution measured for an equivalent random graph with the group found by maximizing modularity.}
\label{fig:bench}
\end{center}
\end{figure}
prominent the disorder of the connections becomes. The scores are shown in Fig.~\ref{fig:bench}a as a function of $\langle k^{ext} \rangle$. Both are able to detect the increasing disorder. Although, as expected, the $\mc$-score is more conservative than the $\mb$-score raising for earlier values of $\langle k^{ext} \rangle$ and so claiming that the group could be found in random graphs before. The (green) continuous curve in the figure represents a numerical estimation of the ideal function that we want to approximate with the scores. Before explaining how it is obtained, we need to describe the second panel of the figure. The distribution for the internal number of connections of $\mc$ is displayed for the benchmarks at different $\langle k^{ext} \rangle$ as well as for equivalent randomized graphs in Fig.~\ref{fig:bench}b. The randomized graphs are obtained by reshuffling the connections of the benchmark networks and the groups of $32$ nodes in them are found by modularity maximization. The curves for the benchmarks start far away in the area of high $m^{int}_\mc$ when $\langle k^{ext} \rangle$ is low. As $\langle k^{ext} \rangle$ increases, they move towards the left and at a certain point, close to $\langle k^{ext} \rangle \approx 8$, cross under the distribution for the randomized graphs. This point marks the end for the significance of the community. Similar (or better) groups could be found in a random graph by a clustering algorithm.  The continuous curve in Fig.~\ref{fig:bench}a is obtained by simulating this process. For each value of $\langle k^{ext} \rangle$, a set of instances of the benchmark are generated. $m^{int}_\mc$ is measured for each of them, and the green curve is calculated averaging the probability of the value $m^{int}_\mc$ or a higher one (cumulative distribution) in the random graph curve of Fig.~\ref{fig:bench}b.
The good agreement of this curve with the $\mb$-score proves that, despite all the approximations, the $\mb$-score is a good measure of cluster significance.

\begin{figure}
\begin{center}
\includegraphics[width=8.6cm]{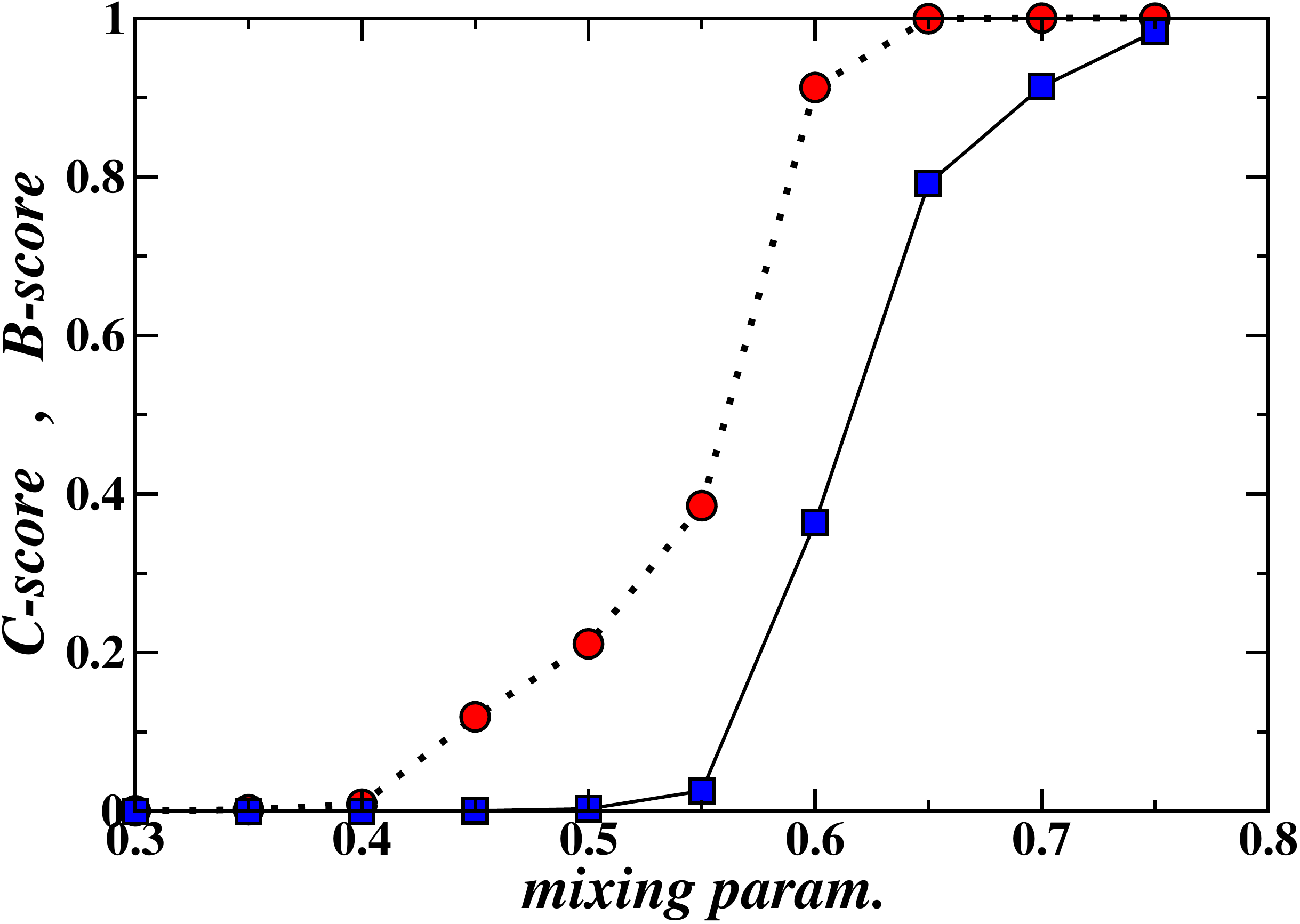}
\caption{(Color online) $\mc$-score (red circles) and $\mb$-score (blue squares) calculated for communities in LFR (heterogeneous) benchmarks. The scores are displayed as a function of the mixing parameter, $k^{ext}/k$. The benchmark networks size is $N = 1000$, the $\langle k \rangle = 15$ with a degree sequence exponent of $\gamma = -2$ and a size of the community $n_\mc = 50$.\label{fig:LFR}}
\end{center}
\end{figure}  

As a final test on benchmarks, we have  evaluated the scores performance on the benchmark proposed by Lancichinetti {\it et al} (LFR) in Ref.~\cite{lanci08}. This technique to generate graphs with planted community structure is a generalization of Girvan and Newman's method to networks with heterogeneous group size and degree distribution. As before, the nodes have $k^{int}$ connections within its own group and $k^{ext} = k -k^{int}$ edges linking elsewhere. The mixing parameter $k^{ext}/k$ indicates the "strength" of the communities. The scores shows a great ability in characterizing the modular structure of the benchmark as we increase the mixing parameter as can be seen in Figure~\ref{fig:LFR}. Due to the absence of fluctuations all the
communities are well defined until each node shares almost half of its connections with nodes of its group, while the groups become less defined for larger values of the mixing parameter. When about the $60\%$ of the links connect with nodes outside the {\it a priori} established groups, the communities become equivalent to those found in random graphs.

\section{Exploring the interior of a community}

\begin{figure}
\begin{center}
\includegraphics[width=8.5cm]{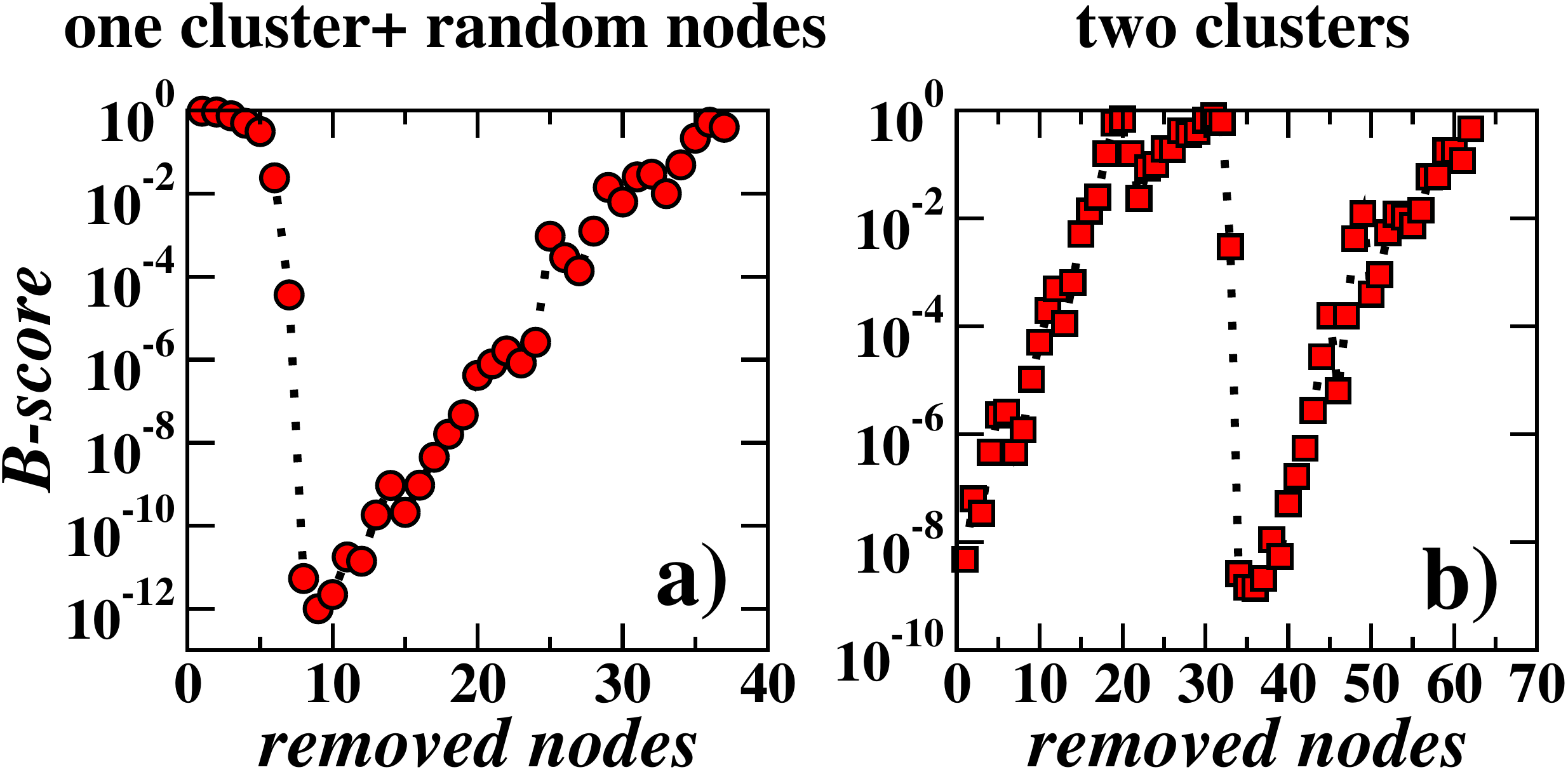}
\end{center}
\caption{(Color online) Iterative application of the $\mb$-score in order to detect the presence of an internal organization in groups. At each stage, we remove the worst node of the community and compute the $\mb$-score for the remaining group. We consider two examples: a) a well defined cluster with the addition of a few random nodes; and b) a group composed by the union of two (good) communities.}
\label{fig:core}
\end{figure}

An interesting application of the scores is the exploration of the internal  structure of groups.
One could decide to remove the worst node from the community as we did to measure the $\mb$-score and recompute the scores for the remaining group. The operation can be repeated iteratively as long as there are nodes remaining in the group. Interestingly, this process is able to identify the presence of internal structure in groups of vertices if the original community displays internal modularity.  Figure~\ref{fig:core} shows two examples of the described operation. The $\mb$-score is plotted as a function of the number of removed nodes. We consider two different examples: a well defined cluster (generated with the LFR benchmark) plus some randomly added nodes (Figure~\ref{fig:core}a); and a group composed of two clusters connected via few random links (Figure~\ref{fig:core}b). The iterative procedure is able to detect and set out the randomly added nodes (Figure~\ref{fig:core}a), and also to find the deeper internal structure inside the two-elements cluster (Figure~\ref{fig:core}b).   

This procedure also allows us to define more detailed measures for the quality of a community. We can search for deeper and deeper cores in the community that we will call {\it $\mc$-$q$ or $\mb$-$q$ core}. Fixed a level of significance $q$, the $\mc$-$q$ (or $\mb$-$q$) core corresponds to the largest sub-group of a community with $\mc$-score ($\mb$-score) lower than $q$. In practical applications, a reasonable value of $q$ is $5\%$. As we will see next, this concept turns out to be a useful tool to characterize communities in real networks. In the case of the benchmarks, the average sizes of the $\mc$-$q$cores obtained for the GN-like networks at $q=5\%$ are close to $32$ up to $\langle k^{ext} \rangle =8$. At this level of disorder, some nodes stop being significant for the planted communities and therefore come excluded from the $q$-core. For higher disorder levels, the cores further reduce until eventually vanish.

 \section{Empirical networks}

\begin{table}
\begin{center}
\begin{tabular}{l|c|c|c|c|c}
\multicolumn{6}{c}{Original partitions} \\ 
\hline
Network & $n_\mc$ & $\mc$-score & $\mb$-score & $\mc$-$5\%$core & $\mb$-$5\%$core\\
\hline
\hline
Karate club~\cite{zachary77}
 & 16 & 0.962 & 0.005 & 0 & 16 \\
$\,$ & 18 & 0.989 & 0.004 & 0 & 18 \\
\cline{1-6}
Karate club & 16 & 0.053 & $10^{-8}$ & 11 & 16 \\
weighted~\cite{zachary77} & 18 & 0.477 & $10^{-9}$ & 17 & 18 \\
\cline{1-6}
$\,$ & 9 & $10^{-10}$ & $10^{-14}$ & 9 & 9 \\
$\,$ & 8 & $10^{-8}$ & $10^{-11}$ & 8 & 8 \\
$\,$ & 11 & $10^{-9}$ & $10^{-12}$ & 11 & 11 \\
$\,$ & 12 & $10^{-9}$ & $10^{-12}$ & 12 & 12 \\
$\,$ & 10 & 1 & 0.941 & 9 & 9 \\
College & 5 & 1 & 0.854 & 0 & 0 \\
football~\cite{girvan02} & 13 & $10^{-8}$ & $10^{-10}$ & 13 & 13 \\
$\,$ & 8 & $10^{-8}$ & $10^{-10}$ & 8 & 8 \\
$\,$ & 10 & $10^{-10}$ & $10^{-13}$ & 10 & 10 \\
$\,$ & 12 & $10^{-10}$ & $10^{-13}$ & 12 & 12 \\
$\,$ & 7 & 0.937 & 0.407 & 4 & 4 \\
$\,$ & 10 & 1 & 0.969 & 8 & 8 \\
\hline
\hline
\end{tabular}
\end{center}
\caption{Analysis of real networks with known community structure. For each network the table reports, from left to right, the name of the network, the size of its communities $n_\mc$, the $\mc$-score , the $\mb$-score, the size of the $\mc$-$5\%$core and the size of the $\mb$-$5\%$core.}
\label{tab:1}
\end{table}

\begin{figure*}
\begin{center}
\includegraphics[width=\textwidth]{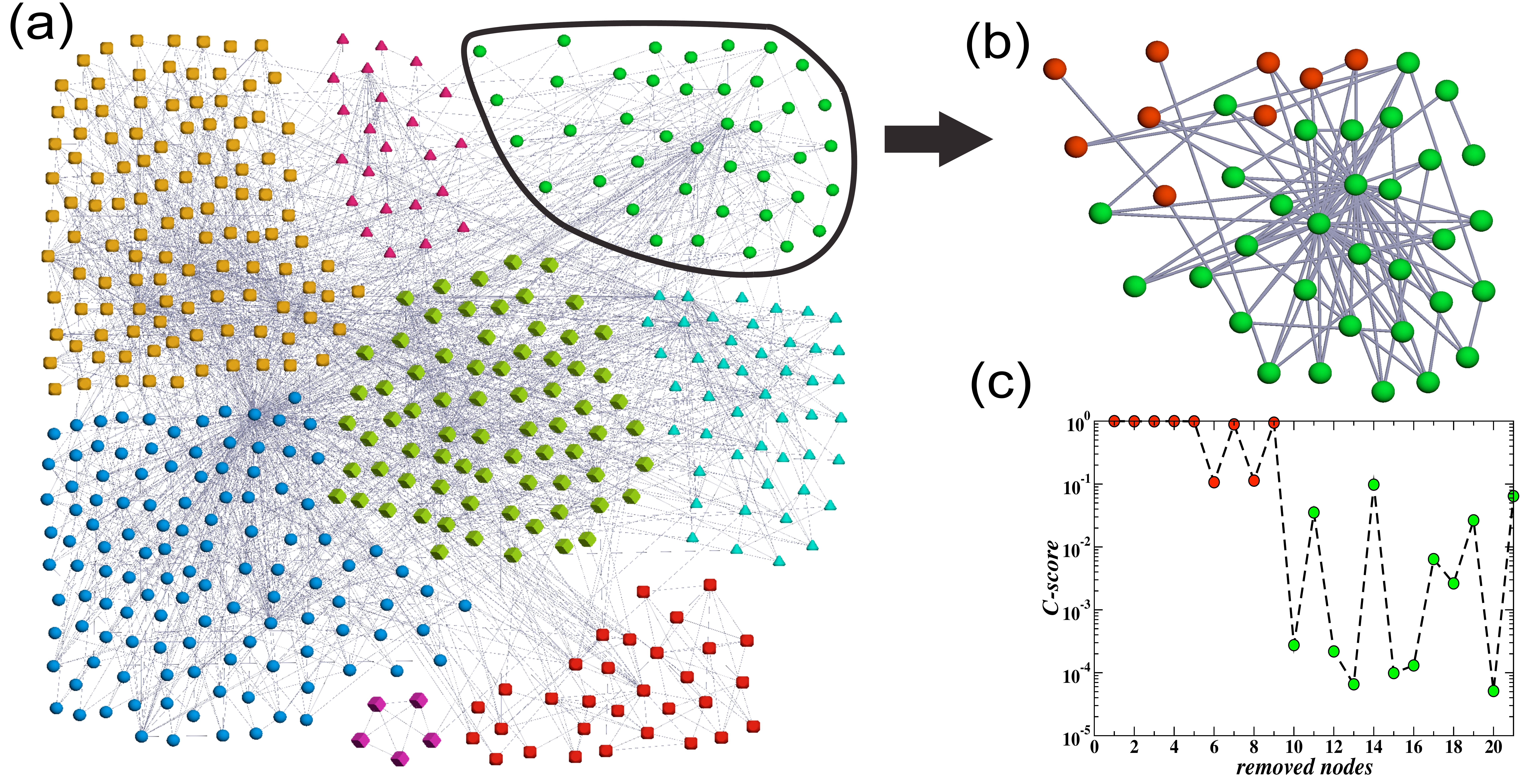}
\caption{(Color online) Community structure for the {\it C.~Elegans} metabolic network~\cite{duch05} obtained by modularity optimization. In (a), an overview of the graph partition is shown. In (b), we display a zoom of a single community depicting in red the nodes that are not significant group members. And (c), the $\mc$-$q$core analysis of the community.\label{realchain} }
\end{center}
\end{figure*}

We show now the utility and versatility of our method for the statistical evaluation of communities in real networks. An exhaustive study of the networks with modular structure in the literature has been performed, the following are only a few examples. We report results on social networks such as the Zachary karate club~\cite{zachary77} or the one extracted for the characters of the novel Les Miserables~\cite{lesmis} or for biological networks such as the {\it C.~Elegans} metabolic network~\cite{duch05}. In two cases, the Zachary club and the college football networks, the structure of the groups is {\it a priori} known. In the Zachary club because the network split in two separate groups due to internal dissensions, and for the college football because the conference in which the teams play is a given data. It is also important to note that some of these networks as, for instance, the Zachary club or the {\it C.~Elegans} metabolic network are weighted graphs for which the weights of the links are equivalent to multiple connections. We have analyzed both the weighted and unweighted versions and report both results in the case of the Zachary club. The  evaluation of the groups for the {\it a priori} known communities is summarized in Table~\ref{tab:1}. While the results for the communities obtained maximizing the  modularity with a simulated annealing technique are displayed in Table~\ref{tab:2}. 
There are some general observations valid for all networks. The $\mc$-score is often able to discriminate good communities, although sometimes a more sophisticated approach as the $\mb$-score is needed. There are also a few cases in which the $\mb$-score reverts the judge based on the $\mc$-score, meaning that a deeper analysis of the communities was required. An example of this type is for instance the Zachary club 2-partition. However, when the original graph with the weight information is considered its communities become more significant. This seems to apply also to the other weighted graphs, showing that there is a connection between clustering structure and weight location in these networks. We also show the sizes of the $5\%$-cores of each community in the Tables as well as detailed analysis of one of the communities of the   {\it C.~Elegans}  metabolic network in Fig.~\ref{realchain}. 

\begin{table}
\begin{center}
\begin{tabular}{l|c|c|c|c|c}
\multicolumn{6}{c}{Maximal modularity partitions} \\ \hline
Network & $n_\mc$ & $\mc$-score & $\mb$-score & $\mc$-$5\%$core & $\mb$-$5\%$core\\
\hline
\hline
$\,$ & 11 & 0.987 & 0.029 & 0 & 11 \\
Karate   & 12 & 0.999 & 0.092 & 0 & 10 \\
club~\cite{zachary77} & 5 & 0.096 & 0.017 & 0 & 5 \\
$\,$ & 6 & 0.996 & 0.505 & 0 & 0 \\
\cline{1-6}
$\,$ & 14 & $10^{-4}$ & $10^{-4}$ & 14 & 14 \\
$\,$ & 10 & 0.005 & 0.003 & 10 & 10 \\
$\,$ & 11 & $10^{-9}$ & $10^{-12}$ & 11 & 11 \\
$\,$ & 15 & 0.088 & 0.001 & 14 & 15 \\
College & 12 & $10^{-9}$ & $10^{-12}$ & 12 & 12 \\
football~\cite{girvan02} & 9 & $10^{-10}$ & $10^{-13}$ & 9 & 9 \\
$\,$ & 10 & $10^{-10}$ & $10^{-13}$ & 10 & 10 \\
$\,$ & 9 & $10^{-6}$ & $10^{-8}$ & 9 & 9 \\
$\,$ & 16 & 0.126 & 0.151 & 12 & 12 \\
$\,$ & 9 & $10^{-10}$ & $10^{-14}$ & 9 & 9 \\
\cline{1-6}
$\,$ & 11 & 0.241 & $10^{-4}$ & 10 & 11 \\
$\,$ & 17 & 0.954 & 0.455 & 15 & 16 \\
Les Miserables& 22 & 0.999 & 0.007 & 9 & 22 \\
~\cite{lesmis}& 10 & 0.868 & $10^{-6}$ & 3 & 10 \\
$\,$ & 11 & 0.999 & $10^{-4}$ & 9 & 11 \\
$\,$ & 6 & $10^{-9}$ & 0 & 6 & 6 \\
\cline{1-6}
$\,$ & 41 (Green) & 1 & 0.832 & 32 & 38 \\
$\,$ & 114 & 1 & 0.001 & 87 & 114 \\
$\,$ & 47 & 0.999 & 0.583 & 35 & 42 \\
{\it C.~Elegans} & 5 & $10^{-6}$ & $10^{-14}$ & 5 & 5 \\
~\cite{duch05}& 23 & 1 & 0.704 & 21 & 21 \\
$\,$ & 73 & 1 & $10^{-10}$ & 67 & 73 \\
$\,$ & 119 & 1 & 0.165 & 50 & 75 \\
$\,$ & 31 & 1 & 0.865 & 25 & 25 \\
\hline
\hline
\end{tabular}
\end{center}
\caption{Analysis of the community structure of several real networks via modularity maximization. For each network the table reports, from left to right, the name of the network, the size of its communities $n_\mc$, the $\mc$-score , the $\mb$-score, the size of the $\mc$-$5\%$core and the size of the $\mb$-$5\%$core. The community highlighted in Figure~\ref{realchain} is marked as (Green) in the table text.}
\label{tab:2}
\end{table}

\section{Conclusion and discussion}

Finding structure in graphs has direct implications for the study of several empirical disciplines as well as for a general understanding of the phenomena behind the evolution of the systems in which such structures raise. Communities are the most direct and easy-to-envisage example of network structures. This concept is a direct heir of the intuitive idea of closer groups when considering social networks. As such, it has had a long history with a good number of algorithms proposed to detect communities in graphs. There are however two important issues missing in the literature. A firm mathematical definition of what a community means and a clear way to determine which of the outputs of the community detection algorithms are really significant. 

In this work, we have focused on the second question with the hope of giving even if partially a hint of where the answer to the first one can lay. 
A new measure able to statistically quantify the meaning of a single community in networks has been introduced. This measure, called $\mc$-score, represents the probability of occurrence of a group with the same properties (i.e., same number of nodes, nodes with the same degree sequence and same internal connections) under the following hypothesis: (i) nodes in the network are randomly connected; (ii) the group is chosen, among all possible groups with the same properties, because is the one which maximizes the density of internal connections. The first hypothesis is a natural assumption and a null model where links are randomly placed is very often used as term of comparison for the determination of correlations or other topological properties in networks. The latter one comes out from the common knowledge which prescribes communities as groups with high intra-connectivity. Thanks to the theory of Extreme Statistics, we approximate the values of the $\mc$-score in the case in which our hypothesis hold. We have tested the performances of the $\mc$-score on several networks, ranging from random graphs to artificial networks with controlled community structure, or to real networks with unknown internal organization. In all cases, we have been able to find good results. The method ability of evaluating one community at a time allows to detect situations in which only some of the communities of the graph are meaningful while the rest of the groups are equivalent to random fluctuations. This approach is also flexible enough to deal with overlapping groups that share nodes between them, providing a separate evaluation for each cluster. Two further refinements of the $\mc$-score have been also introduced. One with the aim of exploring the internal structure of the communities, the $q$-core, and another, the $\mb$-score, with the intention of evaluating a community significance based on a group of nodes instead of on the worst node of the cluster. The computational complexity of the evaluation of the $\mb$- and $\mc$-scores grows quadratically and linearly with the community size, respectively. These tools constitute a set of statistical measures for a thorough evaluation of single communities, avoiding thus the blind acceptance of the output of clustering algorithms.

The software to calculate the $\mc$-score and $\mb$-score of communities is available at~{\tt http://filrad.homelinux.org/cscore}.

\begin{acknowledgments}
The authors would like to warmly thank M. Mungan for his contribution in the discussion that led to Eq.~(\ref{muht}) and H. Nagaraja for his assistance with the Order Statistics distributions involved in the definition of the $\mb$-score. Additionally we thank A. Flammini and S. Fortunato for critical reading of the manuscript and useful suggestions. AL and JJR are funded by the EU Commission projects 238597-ICTeCollective and 233847-Dynanets, respectively.
\end{acknowledgments}

\begin{appendix}
\section{The distribution $Pr\left( < S_{t} \left| \mc_{t}, \mb_{t}, r_{w_{t+1}} \right. \right)$}
\label{app}
In section~\ref{bb}, we have outlined how to compute the $\mb$-score of a community. The iterative procedure makes use of the probability $Pr\left( < S_t \left| \mc_t, \mb_t, r_{w_{t+1}} \right. \right)$ which has yet to be described.  During the procedure for the computation of the $\mb$-score, the size of the border
is increased by one at each stage. At step $t$, the border $\mb_t$ is composed of the $t$ nodes which, on the basis of their internal degrees, are less likely to belong to $\mc_{t-1}$.
We have therefore a sequence  of scores, $r_{w_1} \geq r_{w_2} \geq \ldots \geq r_{w_t}$, for the $t$ worst nodes.  The score of the worst node, namely $w_{t+1}$, still inside $\mc_t$ represents a lower bound for the sequence, since by definition we should have that $r_{w_{t}}\geq r_{w_{t+1}}$.
Instead of trying to obtain the probability for the full sequence, we can simplify our problem and consider the sequence sum $S_t= \sum_{i=1}^{t} \, r_{w_i}$. Finding the distribution of $S_t$ can be formulated as calculating the probability that, given a sequence of $N-n_{\mc_t}$ i.i.d. random variables [we indicate by $n_{\mathcal{F}}$ the size of a set $\mathcal{F}$], the sum of the $t$ largest variables is less than $S_t$.  
The solution for this problem can be found in ~\cite{david03,dempster68}. The cumulative probability distribution is given by the expression

\begin{equation}
\begin{array}{l}
Pr\left( < S_t  \left| \mc_t, \mb_t, r_{w_{t+1}} \right. \right) = 
\\
 1 -  \sum_{j=1}^{\theta_t} (-1)^{j+1} \frac{(n_{\mb_t}+1-j-\xi_t)^{N-n_{\mc_t}-1}}{(n_{\mb_t}+1-j)^{N-n_{\mc}}(n_{\mb_t}-j)!(j-1)!} 
\end{array}
\;\; ,
\label{eq:app1}
\end{equation}

where $\theta_t= \textrm{Integer-Value}\lfloor n_{\mb_t}+1-\xi_t \rfloor$ and $\xi_t=(S_t-n_{\mb_t} w_{t+1})/(1-w_{t+1})$. Note that Eq.~(\ref{eq:app1}) is valid under the assumption of independent variables, which is justifiable to some extent in the case of random networks.

\end{appendix}

\end{document}